\documentclass[prd,twocolumn,showpacs,preprintnumbers,floats,floatfix,apsrev,amsmath,amsfonts,nofootinbib,superscriptaddress]{revtex4-1}
\usepackage{color}
\usepackage{graphicx}
\usepackage{dcolumn}
\usepackage{natbib}
\usepackage{hyperref}
\usepackage{hypernat}
\usepackage{soul}
\newtheorem{thm}{Theorem}

\newcommand{\dd}{{\rm d}}
\newcommand{\be}{\begin{equation}}
\newcommand{\en}{\end{equation}}
\newcommand{\bea}{\begin{eqnarray}}
\newcommand{\ena}{\end{eqnarray}}
\newcommand{\bean}{\begin{eqnarray*}}
\newcommand{\enan}{\end{eqnarray*}}
\newcommand{\p}{\partial}

\begin{document}
\title{Entropy bounds and nonlinear electrodynamics}

\author{F.~T.~Falciano}\email{ftovar@cbpf.br}
\affiliation{CBPF - Brazilian Center for Research in Physics, Xavier Sigaud st. 150,	zip 22290-180, Rio de Janeiro, RJ, Brazil.}
\affiliation{PPGCosmo, CCE - Federal University of Esp\'\i rito Santo, zip 29075-910, Vit\'oria, ES, Brazil.}

\author{M.~L.~Pe\~{n}afiel}\email{mpenafiel@cbpf.br}
\affiliation{CBPF - Brazilian Center for Research in Physics, Xavier Sigaud st. 150,	zip 22290-180, Rio de Janeiro, RJ, Brazil.}

\author{Santiago Esteban Perez Bergliaffa}\email{sepbergliaffa@gmail.com}
\affiliation{Departamento de F\'{\i}sica Te\'orica, Instituto de F\'{\i}sica, Universidade do Estado de Rio de Janeiro, CEP 20550-013, Rio de Janeiro, Brasil.}

\date{\today}
\begin{abstract}
Bekenstein's inequality sets a bound on the entropy of a charged macroscopic body. Such a bound is understood as a universal relation between physical quantities and fundamental constants of nature that should be valid for any physical system. We reanalyze the steps that lead to this entropy bound considering a charged object in conformity to Born-Infeld electrodynamics and show that
the bound
depends of the underlying theory used to describe the physical system. Our result shows that the nonlinear contribution to the electrostatic self-energy causes a raise in the entropy bound. As an intermediate step to obtain this result, we exhibit a general way to calculate the form of the electric field for a given nonlinear electrodynamics in Schwarzschild spacetime.
 \end{abstract}


\maketitle

\section{Introduction}

It follows from black hole mechanics that the entropy of a black hole  (BH) can be identified with its area ~\cite{Bekenstein1972,Wald1984}. The generalized second law~(GSL) states~\cite{Bekenstein1974} that the sum of the entropy of the BH and that of ordinary matter fields outside the BH can never decrease. Considering the infall of an object with radius $\mathcal{R}$ and energy $\mathcal{E}$, Bekenstein first proposed~\cite{Bek81} a universal bound of the entropy of a macroscopic object by
\be \label{Ineq:Bek1}
 S\le{\frac{2\pi k_B}{\hbar c}}\mathcal{ER}\ .
\en

Despite having been derived purely by gravitational considerations, a series of works confirmed the above inequality in a variety of physical situations~\cite{Bek11989,Bekenstein1987,Casini2008,Blanco2013,Myers2012}, 
a fact which
give support to the 
universality of bound \eqref{Ineq:Bek1}. Since Bekenstein's original idea, some generalizations have been proposed to include charge~\cite{Zaslavskii1992,Hod1999} and angular momentum~\cite{Hod99}. The most general bound  reads~\cite{Bekenstein1999}
\be \label{eq:BekBound}
 S\le\frac{2\pi k_B}{\hbar c}\left(\sqrt{\left(\mathcal{ER}\right)^2-c^2J^2}-\frac{Q^2}{2}\right)\ ,
\en
where $\mathcal{R}$ is the radius of the minimum sphere that encloses the system, $J$ the angular momentum, $Q$ the charge\footnote{Depending on the system of units used, a factor $4\pi$ dividing the charge term may be present.} and $c$, $k_B$ and $\hbar$ are respectively the speed of light, Boltzmann and Planck constant\footnote{We use electrostatic units such that Coulomb constant equals 1, hence charge squared has dimensions of energy times length.}. Inequality \eqref{eq:BekBound} is trivially satisfied for non-relativistic systems, and the Kerr-Newman black hole saturates it. Hence, such BH are viewed as the most entropic objects that can be characterized with these three parameters. This is a remarkable result constructed within BH thermodynamics that motivates the interpretation of the upper limit on the entropy given by
(\ref{eq:BekBound})
as a universal relation between physical quantities and fundamental constants of nature that should be valid for every physical system.

Starting from the fact that the entropy is always non-negative, Dain~\cite{Dain2015} derived three subsidiary inequalities relating the size, charge, angular momentum, and energy as direct consequences of \eqref{eq:BekBound} and proved that they hold for any field configuration obeying Maxwell electrodynamics in flat spacetime. However, it can be shown~\cite{Penafiel2017} that nonlinear electrodynamics (NLED) easily violates the inequalities presented in~\cite{Dain2015}. Thus, one must recognize that Bekenstein's inequalities might be theory-dependent.

A minimum requirement for a viable NLED is to recover Maxwell electrodynamics in the appropriate limit to satisfy experimental constraints. Additionally, physical arguments based on causality and unitarity restrict the form of NLED Lagrangians. Among the many nonlinear theories for the electromagnetic field, Born-Infeld (BI) electrodynamics emerges as an interesting alternative to Maxwell's theory. Originally, M. Born and L. Infeld ~\cite{Born1933,Born410,Born425} proposed the nonlinear modification as a means to avoid the classical divergences present in the linear theory. BI can be obtained as the low-energy regime of string theory~\cite{Fradkin1985} and it is unique in the sense that it is the only NLED without birefringence \cite{Boillat1970,Plebanski1970}. 

Bekenstein's inequality was derived via a {\emph{gedanken}} experiment where a spherically symmetric charged particle (its field obeying Maxwell's theory) is slowly lowered radially to  point as close as possible to the horizon and then dropped into the BH. The bound on the entropy appears as a consistency condition related to the change of the BH area. Here we reanalyze the same \emph{gedanken} experiment but assuming that the field of the charged particle obeys BI instead of Maxwell's electrodynamics. In order to describe this process we first need to obtain the BI electric field in the vicinity of a Schwarzschild BH. We shall show that, given the symmetry of the problem, it is possible to find an expression for the electrostatic field of an arbitrary NLED in Schwarzschild spacetime (theorem~\ref{thmNLEDSBH} of sec.~\ref{sec:3}).

In the present work, we calculate the minimum change in the area of the BH due to the absorption of a BI charged body and obtain the corresponding entropy bound. 
Since the contribution to the change in the area comes from the electrostatic
self-energy of the body, it should 
depend of
$q^2$ (which has dimensions of (length)$^2$) and 
$\frac{\mathcal R}{\lambda_{\rm BI}}$, where $\mathcal R $ is the size of the body, and $\lambda_{\rm BI}$ a characteristic length associated to BI theory.
We shall see that our calculation yields precisely such a combination, and that the BI nonlinearities raise the entropy upper limit, thus showing that these  inequalities do depend on the underlying dynamical theory being considered.

The paper is organized as follows: in Sec.~\ref{sec:2} we review Linet's solution for a charged particle in the geometry of a Schwarzschild BH. In Sec.~\ref{sec:3} we generalize the result for an arbitrary NLED and in Sec.~\ref{sec:4} we explicitly give the electric field for the case of BI. In Sec.~\ref{sec:5} we analyze the minimum change of the BH area and obtain the entropy bound considering a BI particle. We close in Sec.~\ref{sec:6} with some comments.

\section{Maxwell Electrodynamics in Schwarzschild spacetime} \label{sec:2}

Electromagnetism is a vector gauge theory for the $U(1)$ symmetry group where the Faraday tensor is given in terms of the dynamical vector field as $F_{\mu\nu}=\partial_{\mu}A_{\nu}-\partial_{\nu}A_{\mu}$. Thus, for any electromagnetic theory, we assume the validity of the second pair of Maxwell's equations, i.e. $ \partial_{[\alpha}F_{\mu\nu]}=0$, where the brackets imply total anti-symmetrization in the indices. The dual of the Faraday tensor is given by $\widetilde{F}^{\mu\nu}={1\over2}\eta^{\mu\nu\alpha\beta}F_{\alpha\beta}$ where $\eta^{\mu\nu\alpha\beta}$ is the totally antisymmetric Levi-Civita tensor. The electric and magnetic fields are defined as the projection of the Faraday tensor and its dual along the normalized observer's worldline $v^\mu$, hence
\begin{displaymath}
\left.
\begin{array}{l}
E^\mu=F^{\mu}_{\ \alpha}v^{\alpha}\\
\\
B^{\mu}=\widetilde{F}^{\mu}_{\ \alpha}v^{\alpha}
\end{array}
\right\}\Rightarrow \ E^\mu v_\mu=B^\mu v_\mu=0\ .
\end{displaymath}
There are only two Lorentz, linearly independent, invariants constructed with the Faraday tensor, its dual and the metric, namely,
\begin{align}\label{eq:FG}
F&\equiv{1\over2}F^{\mu\nu}F_{\mu\nu}=E_{\alpha}E^{\alpha}-B_{\alpha}B^\alpha 
\quad ,\\
G&\equiv {1\over2}\widetilde{F}^{\mu\nu}F_{\mu\nu}=2B_{\alpha}E^{\alpha}.
\end{align}

In a generic Riemannian spacetime with metric $g_{\mu\nu}(x)$, Maxwell's equations read
\begin{align}
\partial_\mu\left(\sqrt{-g}F^{\mu\nu}\right)&=4\pi \sqrt{-g}j^\nu \ ,\label{Max1}
\end{align}
where $g$ is the determinant of the metric tensor, and $j^\nu$ is the current density\footnote{This equation is fully covariant due to the property $\partial_\mu\left(\sqrt{-g}A^{\mu\nu}\right)=\sqrt{-g}\, \nabla_\mu A^{\mu\nu}$ for an arbitrary antisymmetric tensor $A_{\mu\nu}$.}.

The properties of the electric field produced by a static source in Schwarzschild spacetime have been analyzed in several papers (see \cite{Cohen1971,Hanni1973,Linet2000,Molnar2001,Copson1928,Linet_1976}). In particular, Copson~\cite{Copson1928} found the solution for the electrostatic potential of a charged particle in the vicinity of a Schwarzschild BH. However, as shown by Linet~\cite{Linet_1976}, Copson's solution needs to be corrected by adding a spherically symmetric term in order to compensate for the presence of an extra unwanted charged particle inside the BH. Next we follow Linet's work~\cite{Linet_1976} and present the general solution of a static point-like charged particle in the vicinity of Schwarzschild BH. The Schwarzschild metric in standard coordinates $(t,R,\theta,\varphi)$ is given by
\be
\dd s^2=\left(1-{r_s}/R\right)c^2\dd t^2-\frac{\dd R^2}{\left(1-{r_s}/{R}\right)}-R^2\dd \Omega^2
\en
where $r_s=2GM/c^2$, with $M$ being the mass of the black-hole and $\dd \Omega^2\equiv
\dd \theta^2+\sin^2\theta \, \dd \varphi^2$. We shall consider only the electrostatic case where $\mathbf{B}=0$ and $\partial_t \mathbf{E}=0$. Thus, the electric field can be written as the gradient of a scalar function, i.e. $\mathbf{E}=-\boldsymbol{\nabla}\phi$. The time component of equation \eqref{Max1} becomes a second order equation for the potential $\phi$, namely
\begin{align}\label{maxstandard}
\Delta\phi-\frac{r_s}{R^3}\frac{\partial}{\partial R}\left(R^2\frac{\partial \phi}{\partial R}\right)=-4\pi \left(1-\frac{r_s}{R}\right)\rho \ ,
\end{align}
where $\Delta$
is the flat spacetime Laplacian operator given by
\[
\Delta\equiv \frac{1}{R^2}\frac{\partial}{\partial R}\left(R^2\frac{\partial }{\partial R}\right)+\frac{1}{R^2}
\left[\frac{1}{\sin \theta}\frac{\partial}{\partial \theta}\left(\sin\theta\frac{\partial }{\partial \theta}\right)
+\frac{1}{\sin^2 \theta}\frac{\partial^2 }{\partial \varphi^2}
\right]\ .
\]

Linet's solution of \eqref{maxstandard} for a charged particle as source is a modification of Copson's solution~\cite{Copson1928}, which is given in isotropic coordinates $(t,r,\theta,\varphi)$. In such coordinates, the Schwarzschild metric has the form 
\be
\dd s^2=\left(\frac{1-{r_s}/{4r}}{1+{r_s}/{4r}}\right)^2c^2\dd t^2-\left(1+{r_s}/{4r}\right)^4\left[\dd r^2+r^2\dd \Omega^2\right]\ .
\en

Note that in isotropic coordinate system the horizon is located at $r_h=r_s/4=GM/2c^2$. The transformation that connects the standard $(t,R,\theta,\varphi)$ and the isotropic $(t,r,\theta,\varphi)$ coordinates systems prompts the following relations:
\begin{align*}
R&=r\left(1+\frac{r_s}{4r}\right)^2\quad , \quad 2r +\frac{r_s}{2}=R+\sqrt{R(R-r_s)}\\
\frac{\partial r}{\partial R}&=\left(1-\frac{r_s^2}{16r^2}\right)^{-1}\ , \quad
1-\frac{r_s}{R}=\left(\frac{1-{r_s}/{4r}}{1+{r_s}/{4r}}\right)^2  \ .
\end{align*}
It is straightforward to show that, in the isotropic coordinate system, equation \eqref{maxstandard} for the potential $\phi\left(r,\theta,\varphi\right)$ reads
\begin{align}\label{maxisotropic}
\Delta\phi+&
\frac{\left(1-{r_s}/{4r}\right)}{\left(1+{r_s}/{4 r}\right)^3}
\frac{\partial}{\partial r}\left(
\frac{\left(1+{r_s}/{4r}\right)^3}{\left(1-{r_s}/{4 r}\right)}
\right)\frac{\partial \phi}{\partial r}\nonumber\\
&=-4\pi \rho \left(1-\frac{r_s^2}{16 r^2}\right)^{2}\ .
\end{align}
Copson obtained a solution $V_c\left(r,\theta\right)$ for this differential equation corresponding to a charge $q$ located at $\left(r=a,\theta=0,\varphi=0\right)$ outside the horizon, {\emph{i.e.}} $a>r_s/4$, which is given by 
\be\label{copsonsol}
V_c(r,\theta)=\frac{Q}{r\left(1+r_s/4r\right)^{2}}
\left(\mu(r,\theta)+\frac{b}{a}\frac{1}{\mu(r,\theta)}\right),
\en
where
\begin{align}
\mu(r,\theta)&=\sqrt{\frac{\left(r-b\right)^2+2br\left(1-\cos\theta\right)}{\left(r-a\right)^2+2ar\left(1-\cos\theta\right)}},
\label{mufunct}\\
Q&=q\left(1+r_s/4a\right)^{-2}\quad , \qquad b\equiv \frac{r_s^2}{16a}\ .
\end{align}
Linet has shown that this solution describes not one but two charged particles: one of charge $q$ at $(a,0,0)$ and another of charge $-Q r_s/2a$ inside the horizon. In order to correct the solution to describe a single particle of charge $q$ at $( a,0,0)$, Linet added a term $V_p(r,\theta)$, arriving at the solution 
\begin{align}
\psi (r,\theta)&=V_c(r,\theta)+V_p(r,\theta)
=\frac{Q}{r\, \mu}\left(\frac{\mu+r_s/4a}{1+r_s/4r}\right)^2 \ , \nonumber\\
&\mbox{where }\ V_p(r,\theta)=\frac{Q r_s/2a}{ r\left(1+r_s/4 r\right)^{2}}\quad .
\label{linetsol}
\end{align}

In the next section we shall show that the expression for the electric displacement $\mathbf{D}$ in the electrostatic case for an arbitrary NLED in Schwarzschild spacetime is solely given in terms of any potential that is solution of~\eqref{maxisotropic}. Therefore, if the NLED allows to invert the electric field in terms of the electric displacement vector, the electric field $\mathbf{E}$ and its potential $\phi(r,\theta)$ can be expressed in terms 
of the potential $\psi(r,\theta)$.

\section{Nonlinear electrodynamics in Schwarzschild spacetime} \label{sec:3}

Nonlinear electrodynamics can be defined as generalizations of Maxwell electrodynamics obtained 
by changing the Larmor Lagrangian density $(\mathcal{L}=-F/8\pi)$ to arbitrary functions of the two invariants $F$ and $G$. Although Maxwell electrodynamics is well-confirmed by experiments,  there are interesting theoretical arguments~\citep{Born1933, Born410, Fradkin1985, Metsaev1987, Tseytlin1997, Gibbons2001, Abalos2015, Euler1936, Delphenich2003} that motivate the  examination of NLED. There are several examples of NLED in the literature~\citep{Gaete2014, Gaete2014a, Hendi2012, Kruglov2015, Kruglov2017, Dunne2004}. The minimal condition is that the theory should recover Maxwell dynamics in the appropriate limit but it is also desirable to satisfy the causality and unitarity conditions~\cite{Shabad2011,Schellstede2016,Goulart2009}. Apart from these conditions
most analyses maintain minimal coupling with matter source, hence the coupling between field and current density is given by the usual combination $j^{\mu}A_\mu$. The dynamics for the Lagrangian $\mathcal{L}(F,G)$ reads
\begin{eqnarray}
\p_{\mu}(\sqrt{-g}E^{\mu\nu})=-4\pi\sqrt{-g}j^{\nu} \quad ,\label{eq:fNLED}
\end{eqnarray}
where $E^{\mu\nu}$ is the excitation tensor, which together with its dual are defined as
\begin{subequations}
\begin{eqnarray}
E^{\mu\nu}&=\frac{\p\mathcal{L}}{\p F_{\mu\nu}}=2\left(\mathcal{L}_FF^{\mu\nu}+\mathcal{L}_G\widetilde{F}^{\mu\nu}\right)\quad ,\label{Eq:def:Emunnu}\\
\widetilde{E}^{\mu\nu}&=\frac12\eta^{\mu\nu\alpha\beta}E_{\alpha\beta}=2\left(\mathcal{L}_F \widetilde{F}^{\mu\nu}-\mathcal{L}_G F^{\mu\nu}\right) \ .\label{Eq:def:tEmunnu}
\end{eqnarray}
\end{subequations}

The notation $\mathcal{L}_{X}$ means derivative of the Lagrangian with respect to $X$. Similarly to the invariants $F$ and $G$, we can define two invariant quantities using $E^{\mu\nu}$ and $\widetilde{E}^{\mu\nu}$, namely
\begin{subequations}
\begin{eqnarray}
P&={1\over2}E^{\mu\nu}E_{\mu\nu}=4\left(\mathcal{L}_{F}^2-\mathcal{L}_{G}^2\right)F+8\mathcal{L}_F\mathcal{L}_{G}G \ ,\\
S&={1\over2}\widetilde{E}^{\mu\nu}E_{\mu\nu}=4\left(\mathcal{L}_{F}^2-\mathcal{L}_{G}^2\right)G-8\mathcal{L}_F\mathcal{L}_{G}F\ .
\end{eqnarray}
\end{subequations}
The decomposition of the excitation tensor follows closely that of the Faraday tensor, namely
\begin{displaymath}
\left.
\begin{array}{l}
D^\mu=-E^{\mu}_{\ \alpha}v^{\alpha}\\
\\
H^{\mu}=-\widetilde{E}^{\mu}_{\ \alpha}v^{\alpha}
\end{array}
\right\}\Rightarrow \ D^\mu v_\mu=H^\mu v_\mu=0\ ,
\end{displaymath}
where $D^\mu$ and $H^\mu$ are respectively the four-dimensional electric displacement and magnetic $H$-field. The equations of motion for the fields can be rewritten in terms of the excitation tensors using a Legendre transformation. The resulting formulation is called P-framework~
\cite{Plebanski1970,Salazar1987,Bronnikov2001} and it is based on the associated Hamiltonian-like density\footnote{The factor 1/2 in the first term appears due to the notation conventions we are using for $F^{\mu\nu}$. This Hamiltonian-like density coincides with the energy density for NLED in flat spacetime.} defined as
\begin{equation} \label{eq:Hden}
\mathcal{H}={1\over2}E_{\mu\nu}F^{\mu\nu}-\mathcal{L}=2\left(\mathcal{L}_{F}F+\mathcal{L}_{G}G\right)-\mathcal{L}\ .
\end{equation}

In order to complete the above Legendre transformation and write $\mathcal{H}$ as a function of $E_{\mu\nu}$ and $\widetilde{E}_{\mu\nu}$ we need to invert \eqref{Eq:def:Emunnu} and write $F^{\mu\nu}$ as a function of the excitation tensor and its dual. If \eqref{Eq:def:Emunnu}-\eqref{Eq:def:tEmunnu} are invertible, we can write 
\begin{equation}
F^{\mu\nu}=
2\frac{\p\mathcal{H}}{\p E_{\mu\nu}}
=2\left(\mathcal{H}_{P}E^{\mu\nu}+\mathcal{H}_{S}\widetilde{E}^{\mu\nu} \right)\quad ,
\end{equation}
where $\mathcal{H}_{P}$ and $\mathcal{H}_{S}$ represent the derivatives of the Hamiltonian density with respect to the invariants $P$ and $S$, respectively. Therefore, the Lagrangian can be written in terms of the Hamiltonian an its derivatives as 
\begin{equation}
\mathcal{L}(P,S)=2\left(\mathcal{H}_{P}P+\mathcal{H}_{S}S\right)-\mathcal{H}\quad .
\end{equation}

Our strategy will be to solve \eqref{eq:fNLED} for the excitation tensor in Schwarzschild spacetime and then use \eqref{Eq:def:Emunnu}-\eqref{Eq:def:tEmunnu} to find the electromagnetic fields. This procedure can be stated as:

\begin{thm}\label{thmNLEDSBH}
The electrostatic potential $\phi(x)$ produced by a charged particle in a generic NLED theory $\mathcal{L}(F,G)$ in Schwarzschild spacetime is entirely specified by the electrostatic potential $\psi(x)$ satisfying Maxwell's electromagnetism in the same background. The displacement vector is curl-free and given by $\mathbf{D}=-\boldsymbol{\nabla}\psi(x)$.
\end{thm}

\textit{Proof.}--- For the pure electrostatic case, namely $\mathbf{B}=0$ and $\partial_t\mathbf{E}=0$, the electric displacement is given solely in terms of the electric field as $\mathbf{D}=-2 \mathcal{L}_F(E)\, \mathbf{E}$, with $\mathbf{E}=-\boldsymbol{\nabla}\phi$. Thus, in Schwarzschild spacetime and using the isotropic coordinate system, it is straightforward to show that \eqref{eq:fNLED} becomes
\begin{align}\label{NLEDisotropic}
\Delta \phi+
\frac{\left(1-{r_s}/{4r}\right)}{\left(1+{r_s}/{4 r}\right)^3}
\frac{\partial}{\partial r}\left(
\frac{\left(1+{r_s}/{4 r}\right)^3}{\left(1-{r_s}/{4 r}\right)}
\right)\frac{\partial \phi}{\partial r}=\nonumber \\
=\frac{2\pi \rho}{\mathcal{L}_F(\boldsymbol{\nabla}\phi)} \left(1-\frac{r_s^2}{16 r^2}\right)^{2}-\frac{1}{\mathcal{L}_F( \boldsymbol{\nabla}\phi)}
\boldsymbol{\nabla}\phi \cdot \boldsymbol{\nabla} \mathcal{L}_F\ .
\end{align}

Let $\psi(x)$ be an auxiliary scalar function defined as the integral along the path with tangent vector $\mathbf{\dd l}$ such that 
\begin{align} \label{def:auxfunc}
 \psi ( x)&= -2 \int  \ \mathcal{L}_F(\boldsymbol{\nabla}\phi) \boldsymbol{\nabla}\phi \cdot \mathbf{\dd l}\nonumber\\
 \Rightarrow \  \boldsymbol{\nabla}\psi (x)&= -2  \mathcal{L}_F(\boldsymbol{\nabla}\phi)
 \boldsymbol{\nabla}\phi\ .
\end{align}

Inserting in \eqref{NLEDisotropic} we find
\begin{align}
\label{NLEDisotropic2}
\Delta \psi & +
\frac{\left(1-{r_s}/{4r}\right)}{\left(1+{r_s}/{4r}\right)^3}
\frac{\partial}{\partial r}\left(
\frac{\left(1+{r_s}/{4 r}\right)^3}{\left(1-{r_s}/{4 r}\right)}\right)\frac{\partial \psi}{\partial r}
= \nonumber \\
& -4\pi \rho\left(1-\frac{r_s^2}{16r^2}\right)^{2} \ ,
\end{align}
which is exactly \eqref{maxisotropic}. Thus, we can identify $\psi (x)$ as the electrostatic potential for Maxwell electrodynamics in Schwarzschild. Furthermore, the excitation tensor $\mathbf{D}=-2 \mathcal{L}_F(E)\, \mathbf{E}$ is minus the gradient of $\psi$ and inverting this relation we find the electric field as a nonlinear function of $\psi(x)$ and its derivatives. \hfill$\square$

\section{Born-Infeld Electrodynamics in a Schwarzschild Black Hole} \label{sec:4}

Theorem \ref{thmNLEDSBH} provides the electrostatic solution for a generic NLED in Schwarzschild spacetime but the physical features depend on the particular NLED theory chosen. In this section we shall particularize to BI electrodynamics. The BI action reads
\begin{equation} \label{eq:LBI}
S=\frac{1}{4\pi}\int \dd^4 x\sqrt{-g} \ \beta^2\left(1-\sqrt{U}\right)\quad ,
\end{equation}
where $\beta $ is a field strength parameter and $U=1+F/\beta^2-G^2/4\beta^4$. In vector notation, the constitutive relations read 
\begin{align}\label{constEqs1}
\mathbf{D}\equiv {1\over{\sqrt{U}}}\left(\mathbf{E}+{\left(\mathbf{E}\cdot\mathbf{B}\right)\over{\beta^2}}\mathbf{B} \right)
\ , \ \mathbf{H}\equiv {1\over{\sqrt{U}}}\left(\mathbf{B}-{\left(\mathbf{E}\cdot\mathbf{B}\right)\over{\beta^2}}\mathbf{E} \right) \ . 
\end{align}
BI electrodynamics is an example of a NLED for which the $P$-framework is well-defined and completely analogous to the $F$-framework. Indeed, defining $V=1-P/\beta^2-S^2/4\beta^4$ one can show that
\begin{align}
S=G\  ,\quad P=F-\frac{F^2+G^2}{U\beta^2} 
\ , \quad
V=\frac{\left(1+{G^2}/{4\beta^4}\right)^2}{U}\quad .
\end{align}
Using the above relations we can invert the constitutive relations \eqref{Eq:def:Emunnu}-\eqref{Eq:def:tEmunnu} and find
\begin{subequations} \label{fftilde}
\begin{eqnarray}
F^{\mu\nu}&= -\frac{1}{\sqrt{V}}\left(
E^{\mu\nu}+\frac{S}{2\beta^2} \widetilde{E}^{\mu\nu} 
 \right)\quad ,\\
\widetilde{F}^{\mu\nu}&= -\frac{1}{\sqrt{V}}\left(
 \widetilde{E}^{\mu\nu}-\frac{S}{2\beta^2} E^{\mu\nu} 
 \right)\quad .
 \end{eqnarray}
\end{subequations}
A straightforward calculation shows that the Hamiltonian for the BI electrodynamics reads $\mathcal{H}=\beta^2\left(\sqrt{V}-1\right)$. Decomposing \eqref{fftilde} in terms of the electric and magnetic fields we have
\begin{align}
\label{constEqs}
\mathbf{E}\equiv {1\over{\sqrt{V}}}\left(\mathbf{D}+{\left(\mathbf{D}\cdot\mathbf{H}\right)\over{\beta^2}}\mathbf{H} \right)
	\ , \ \mathbf{B}\equiv {1\over{\sqrt{V}}}\left(\mathbf{H}-{\left(\mathbf{D}\cdot\mathbf{H}\right)\over{\beta^2}}\mathbf{D} \right) \ . 
\end{align}
In particular, for the electrostatic case, these relations simplify to
\begin{align}\label{constEqsESt}
\mathbf{D}= 
{\mathbf{E} \over{\sqrt{1- |\mathbf{E}|^2 \beta^{-2} }}}  &\quad \mbox{and}&
\mathbf{E}= 
{\mathbf{D} \over{\sqrt{1+ |\mathbf{D}|^2 \beta^{-2} }}} \ . 
\end{align}

Let us consider for a moment the static solution for a point-like particle in Minkowski spacetime. A point-like particle with charge $q$ at rest at $r=a$ is described by $j^\mu=\left(q\delta(r-a),\mathbf{0}\right)$. In electrostatics, the field equations reduce to
\be\label{BIMinkD}
\boldsymbol{\nabla} \cdot \mathbf{D}=4\pi \rho\ \Rightarrow \quad \mathbf{D}=\frac{q}{\left(r-a\right)^3}\left(\mathbf{r}-\mathbf{a}\right)=\frac{\beta}{x^2}\mathbf{\hat{x}}\quad ,
\en 
where the adimensional variable $x$ is defined as $x\equiv \sqrt{\frac{\beta}{q}}\left(r-a\right)$. Using the constitutive equation \eqref{constEqsESt}, the electrostatic field reads
\be\label{BIMinkE}
\mathbf{E}=\frac{\beta}{\sqrt{1+x^4}}\mathbf{\hat{x}}\quad .
\en 
Note that, given the symmetry of the problem, \eqref{BIMinkD}-\eqref{BIMinkE} also describe the field outside a spherically symmetric charged body. The electrostatic potential $\phi_{\rm BI}$ in flat spacetime is given by
\begin{align}
\phi_{\rm BI}(r)&= -\int^{r}_{\infty}{\dd \mathbf{r}\cdot\mathbf{E}}
=\frac{\sqrt{q\beta}}{x} \ {}_2F_1 \left[\frac14,\frac12,\frac54,-\frac1{x^4}\right]
\quad ,\label{BIMinkPot}
\end{align} 
where ${}_2F_1[a,b;c;z]$ is the Gaussian hypergeometric function. We shall need below the following two well-defined limits
\begin{align}
\lim_{x\rightarrow 0}&\quad  {\, {}_2F_1\left[\frac14,\frac12,\frac54,-x^4\right]}=1+\mathcal{O}(x^4)\quad ,\label{Hyperlim0}\\
\lim_{x\rightarrow 0}& \quad \frac1x \, { \, {}_2F_1 \left[\frac14,\frac12,\frac54,-\frac{1}{x^4}\right]}=\frac{\Gamma\left(\frac14\right)\Gamma\left(\frac54\right)}{\sqrt{\pi}}-x+\mathcal{O}(x^5)
\label{Hyperlim1/0}
\end{align}
In particular, the second limit implies that the potential at the position of the particle is finite, and given by 
\be
\phi_{\rm BI}(a)=\frac{\Gamma\left(\frac14\right)\Gamma\left(\frac54\right)}{\sqrt{\pi}}\sqrt{q\beta}\equiv\frac{q}{\lambda_{\rm BI}}\quad .\label{BIMinkPotata}
\en
The last equality defines a characteristic length scale associated with BI electrodynamics, given by $$\lambda_{\rm BI}\equiv \frac{\sqrt{\pi}}{\Gamma\left(\frac14\right)\Gamma\left(\frac54\right)}\sqrt{q/{\beta}}\approx0.54\sqrt{q/\beta}\ .$$ 

The Maxwellian limit is obtained by $\beta\rightarrow\infty$, hence the length $\lambda_{\rm BI}$ can be interpreted as the radius within which the BI nonlinearities become effective. For $r\gg \lambda_{\rm BI}$ Maxwell's electrodynamics is to be recovered.

\begin{figure}[t]
\includegraphics[width=0.47\textwidth]{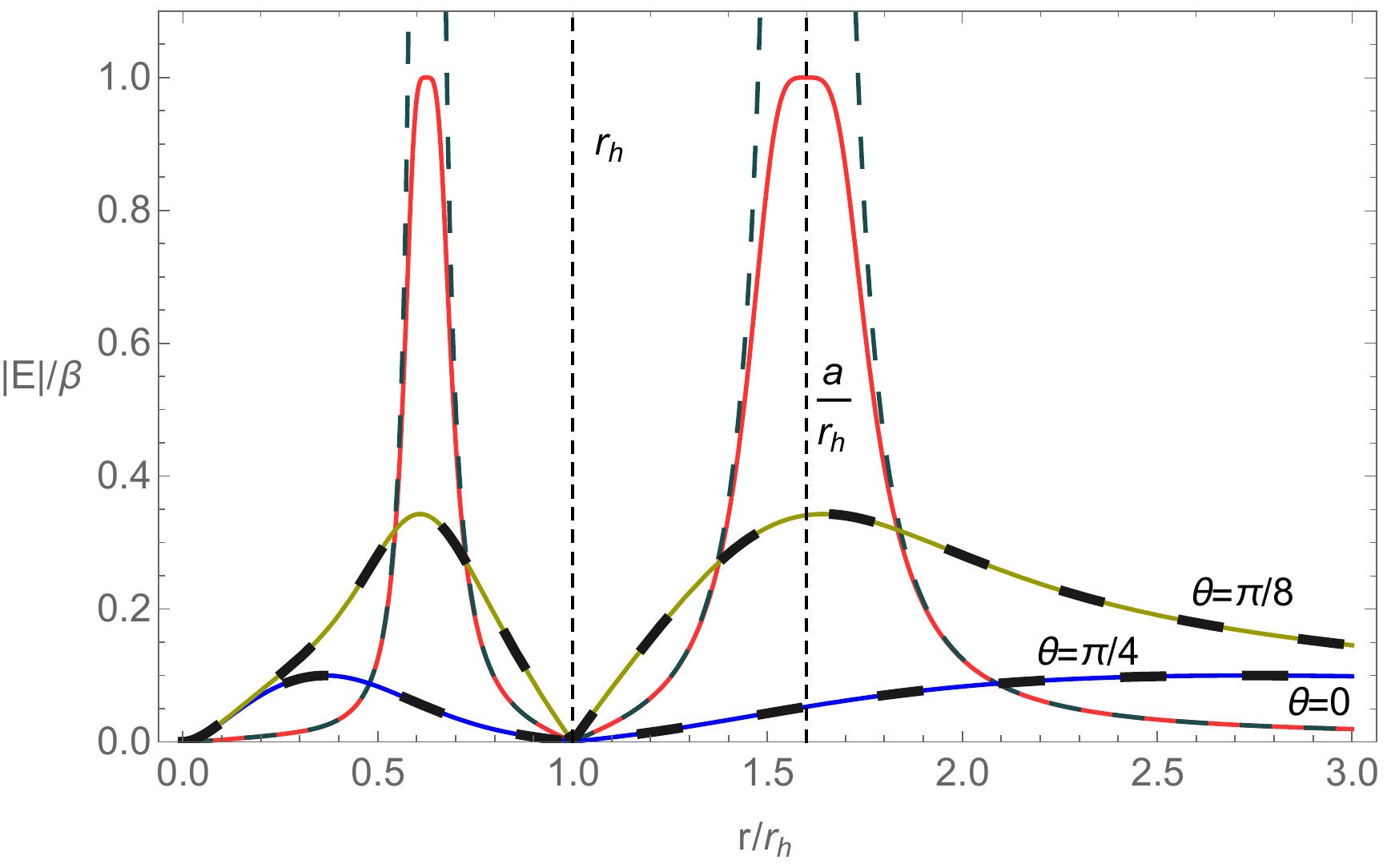}
\caption{Comparison of the  BI solution for the electric field given by the integration of Eqn.\eqref{constitBI2} and Maxwell's electric field for a charged point-like particle outside a Schwarzschild BH. The particle is located at $\left(\theta=0,a=1.6 r_h\right)$ with the horizon $r_h=M/2$ in geometrized units. We set the mass of the BH $M=2$, the particle's charge $q=2\times 10^{-4}$ and the BI parameter $\beta=4\times 10^{-4}$. Solid curves correspond to BI solutions for $\theta=0$, $\theta={\pi}/{4}$, and $\theta={\pi}/{8}$ (the last two rescaled by a factor 10), while dashed lines are their Maxwellian versions.}
\label{fig:BI_Max_EField}
\end{figure}

Let us now turn to curved spacetime. The expressions for the displacement vector and the electric field in Schwarzschild spacetime are straightforwardly given by Theorem~\ref{thmNLEDSBH} of the last section. The electric displacement is the gradient of Linet's potential $(\mathbf{D}=-\boldsymbol{\nabla}\psi)$, hence the BI electric field in a Schwarzschild spacetime reads
\be\label{constitBI2}
\mathbf{E}(r,\theta)=-\boldsymbol{\nabla}\phi
(r,\theta)=-\frac{\boldsymbol{\nabla}\psi(r,\theta)}{\sqrt{1+|\boldsymbol{\nabla}\psi(r,\theta)|^2\beta^{-2}}}\ , 
\en
where $\psi(r,\theta)$ is specified by \eqref{linetsol}. As in the case of the electric field in flat spacetime, the electric field given by \eqref{constitBI2} is finite everywhere, even at the particle position. In Fig.~\ref{fig:BI_Max_EField} we plot the electric field for Maxwell and BI theories in Schwarzschild spacetime as a function of the isotropic radial coordinate for three fixed angles. As expected, far from the particle both solutions coincide. Note also that both fields vanish on the horizon, a fact that is consistent with interpreting the horizon as a conducting surface~\cite{Price1986,MacDonald1985}.

While we could not find an exact expression for the electrostatic potential $\phi(r,\theta)$ using 
Eqn.\eqref{constitBI2}, for our purpose all we need is an expression for $\phi$ near and at the particle. In the limit approaching the particle $\psi(r,\theta)$ diverges, the divergence being driven by the behaviour of $\mu(r,\theta)$ (see \eqref{mufunct}), hence it is Copson's solution $V_c(r,\theta)$ that needs to be carefully analyzed close to the particle while the second part $V_p(r,\theta)$ is regular everywhere (see \eqref{linetsol}). Equivalently, close to the particle, Linet's potential is dominated by Copson's solution, i.e. $\psi(r,\theta)\approx V_c(r,\theta)$ for $r\rightarrow a$ and $\theta\rightarrow 0$.

The nonlinearity of the electric field precludes us to use the superposition principle, but we can still add displacement vector fields $\mathbf{D}=-\boldsymbol{\nabla} \psi$. In other words, 
we can add $\psi$'s
but we cannot add $\phi$'s
to obtain exact solutions to the BI equations. In addition, since Copson's potential $V_c(r,\theta)$ and the extra term $V_p(r,\theta)$ are solutions of Einstein-Maxwell equations in Schwarzschild background, using these potentials in
Eqn.\eqref{constitBI2} automatically gives us two exact solutions $\phi_c(r,\theta)$ and $\phi_p(r,\theta)$ of BI in Schwarzschild background. Note however that $V_p(r,\theta)$ is already regular everywhere and satisfies $|\boldsymbol{\nabla} V_p|<<\beta$, hence for the extra term the BI nonlinearities are uneffective and $\phi_p \approx V_p$.

In order to circumvent the lack of an analytical expression for $\phi$, we shall construct an approximate solution by first decomposing the exact solution for the electric field as a sum of two terms, as follows:
\be\label{aproxBIVcVp}
\mathbf{E}(r,\theta)=-\boldsymbol\nabla\phi \approx-\frac{\boldsymbol{\nabla} V_c}{\sqrt{1+|\boldsymbol{\nabla} V_c|^2\beta^{-2}}}
-\boldsymbol\nabla V_p \ ,
\en
which is equivalent to decomposing the BI electrostatic potential as
\be \label{eq:phiapproximation}
\phi(r,\theta)\approx\phi_c(r,\theta) +V_p(r,\theta) \ .
\en

One can show (see Appendix) that in the limit approaching the particle
\begin{align}
\lim_{r\rightarrow a } \boldsymbol\nabla V_c &\approx   \frac{V_c}{\mu}\boldsymbol\nabla \mu \quad \Rightarrow \quad |\boldsymbol\nabla V_c| \approx  \beta \frac{V_c^2}{\Sigma^2} \ ,
\end{align}
with
\begin{align}
\Sigma^2\equiv  \frac{\beta \, q(1-b/a)}{ \left(1+{r_s}/{4a}\right)^2} \ .
\end{align}
Thus, we can make the following approximations:
\begin{align}\label{phicp}
\boldsymbol{\nabla} \phi_c =
&\frac{\boldsymbol{\nabla} V_c}{\sqrt{1+|\boldsymbol{\nabla} V_c|^2\beta^{-2}}}
\approx \Sigma\frac{\boldsymbol{\nabla} \left(V_c/\Sigma\right)}{\sqrt{1+\left(V_c/\Sigma\right)^4}} \quad  \nonumber\\
\Rightarrow \quad &\phi_c (r,\theta)= -\frac{\Sigma^2}{V_c} \ {}_2F_1 \left[\frac14,\frac12,\frac54,-\left(\frac{\Sigma}{V_c}\right)^4\right]+C_{\infty}
\end{align}
where $C_{\infty}$ is an integration constant. Using the limit given in \eqref{Hyperlim1/0} in \eqref{phicp} we have 
\begin{align}
\lim_{r\rightarrow \infty}
\phi_c (r,\theta)&= C_{\infty}-\Sigma \frac{\Gamma\left(\frac14\right)\Gamma\left(\frac54\right)}{\sqrt{\pi}} +V_c +\mathcal{O}(r^{-5})\quad .
\end{align}
In order to reobtain $\phi\approx\psi$ far away from the particle, the condition $\sqrt{\pi}C_{\infty}=\Sigma\Gamma\left(\frac14\right)\Gamma\left(\frac54\right) $ must be imposed. Thus, the contribution of the Copson solution to the BI electrostatic potential 
close to the particle reads
\begin{align}\label{constitBI5}
\phi_c (r,\theta)=&
\Sigma\left(\frac{\Gamma\left(\frac14\right)\Gamma\left(\frac54\right)}{\sqrt{\pi}} 
-\frac{\Sigma}{V_c} \ {}_2F_1 \left[\frac14,\frac12,\frac54,-\left(\frac{\Sigma}{V_c}\right)^4\right]\right) \  .
\end{align}

\begin{figure}[t]
	\includegraphics[width=0.44\textwidth]{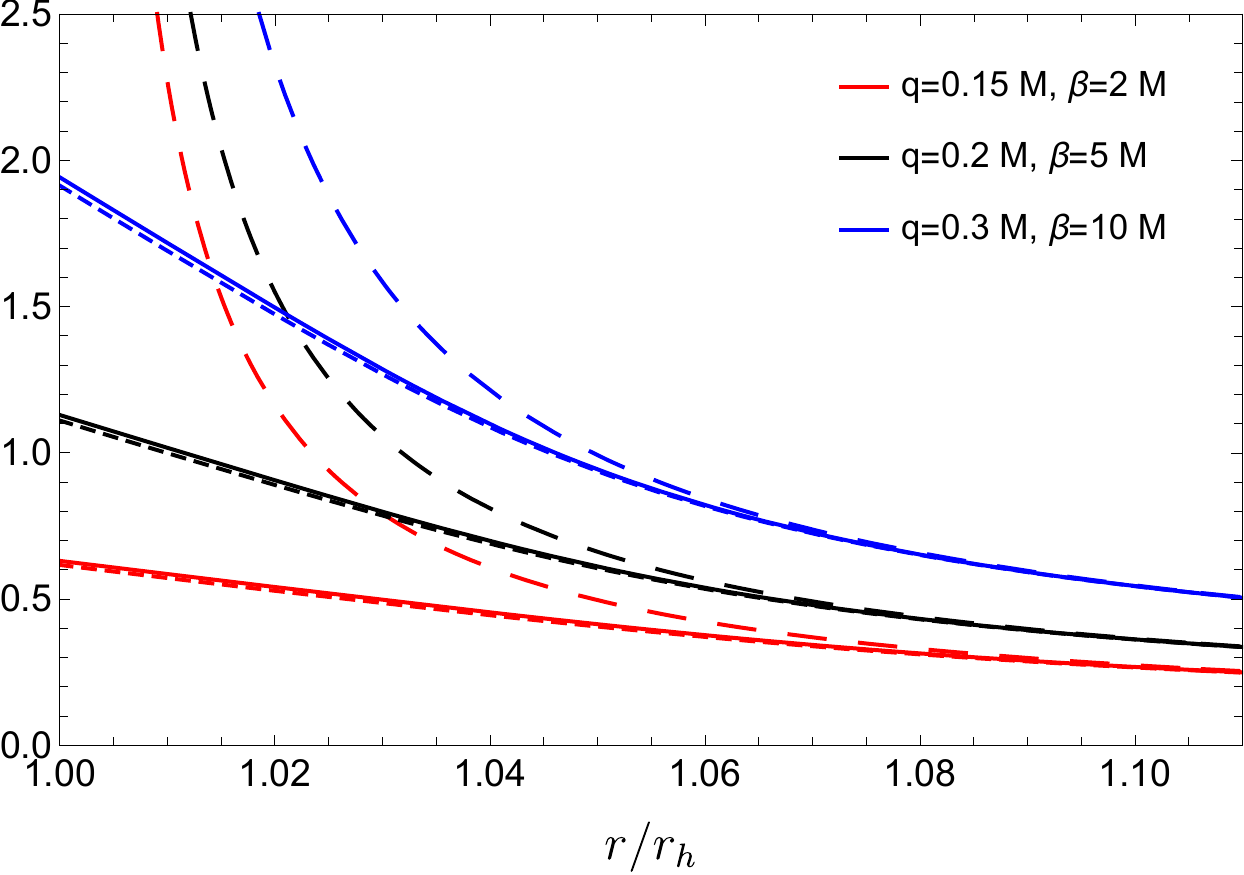}
	\caption{In order to check the validity of our approximations, we plot the numerical solution given by integrating \eqref{constitBI2} and the approximate solution given by \eqref{constitBI5}. The long-dash lines correspond to Linet's potential, solid lines
	represent the solution
	obtained by numerical integration, and short-dash lines correspond to our approximate expression. Far away from the particle all three solutions coincide. We have normalized every quantity in units of the BH mass $M$. The horizon is located at $r_h=M/2$ and we placed the charged particle at $\left(a=2.5r_h,\theta=0\right)$.}
	\label{fig:Valid_Pot_aprox2}
\end{figure}

\begin{figure}[t]
\includegraphics[width=0.47\textwidth]{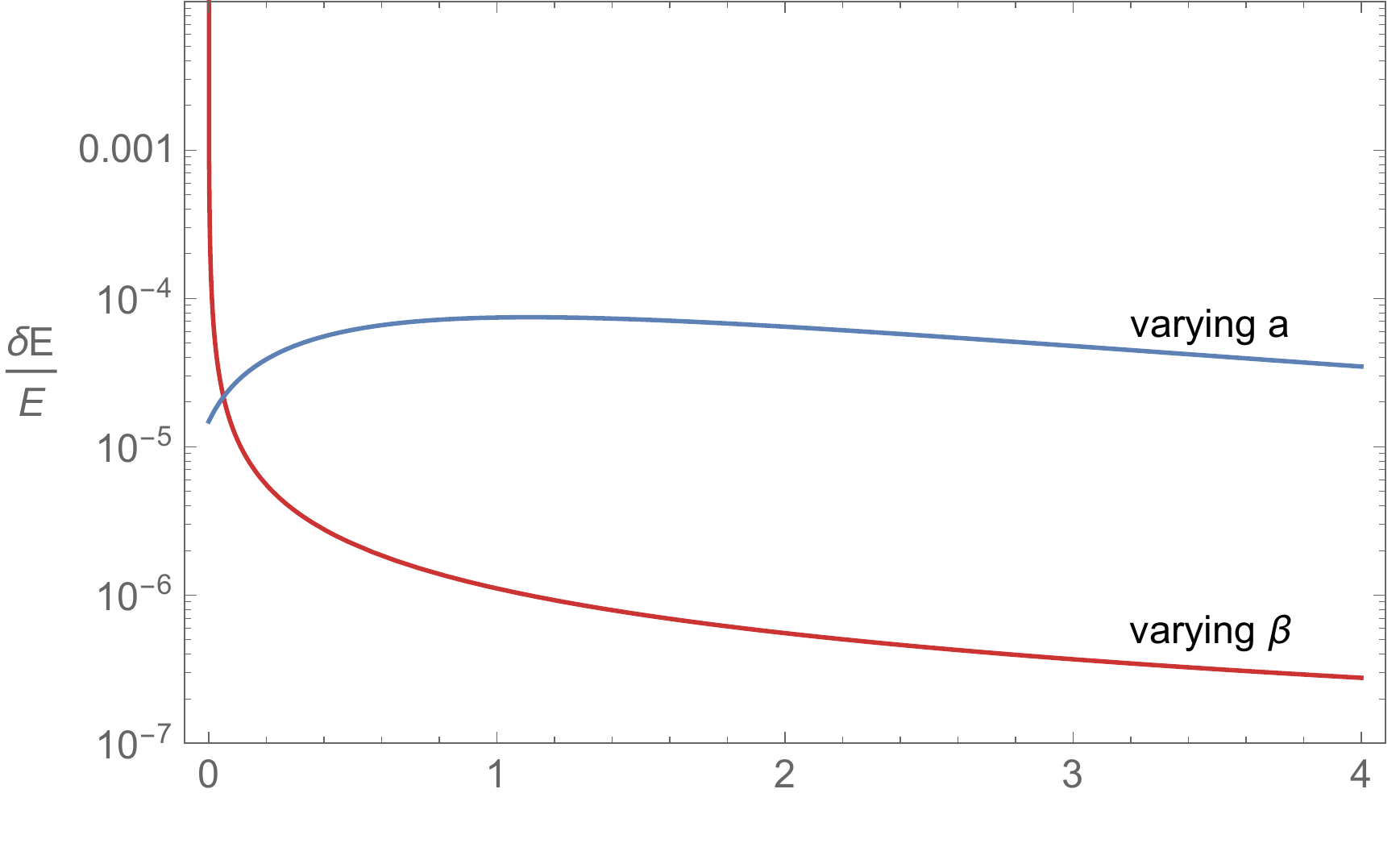}
\caption{Variation of the approximation \eqref{aproxBIVcVp} as a function of the free parameters of the solution. We normalize all parameter by the mass of the BH in geometrized units. We plot the relative error given by the difference of the exact solution \eqref{constitBI2} and~\eqref{aproxBIVcVp} divided by the exact solution. The error decreases rapidly as the BI parameter $\beta$ increases (lower curve). In addition, the approximation is quite independent of the position of the charged particle. 
}\label{fig:AppVarPar}
\end{figure}

Figure~\ref{fig:Valid_Pot_aprox2} compares the exact solution with the approximate solution. The large dashed lines depict the divergent Copson's solution while small dashed and solid lines are respectively the approximate and the exact solutions. Figure~\ref{fig:AppVarPar} displays the fractional difference between the numerical solution obtained by integrating \eqref{constitBI2} to find $\phi_c$ in terms of $V_c$ to the above approximate solution, hence it gives the goodness of the approximation
given in 
Eq.\eqref{constitBI5}
in terms of the dependence on the parameters of the system. For small values of the BI field strength parameter $\beta$ the error drastically decreases. These plots show that close to the particle \eqref{constitBI5} indeed is a good approximation for the electrostatic potential of the BI particle in the geometry of a Schwarzschild BH.

In the limit approaching the charged particle, the term associated to the Copson's potential goes to a constant given by
\begin{equation}
\phi_c(a,0)=\phi_{\rm BI}\frac{\sqrt{\left(1-b/a\right)}}{\left(1+r_s/4{a}\right)}
=\phi_{\rm BI}\sqrt{\frac{\left(1-r_s/4{a}\right)}{\left(1+r_s/4{  a}\right)}}
\end{equation}
where $\phi_{\rm BI}$ is the electrostatic BI potential at the particle's position for flat spacetime (see Eqn. \eqref{BIMinkPotata}).

Analyzing the exact solution, one can show that the gradient of Linet's potential goes to zero at the horizon independently of the particle's position. Therefore the BH's horizon works as a conducting sphere~\cite{Price1986,MacDonald1985} that forces the electric field to go to zero. A charged particle close to a conductive sphere polarizes its surface, hence, the presence of the charged particle polarizes the BH. This polarization can be described by an image particle inside the BH. As a result, we need to include the contribution of the image particle for the total electrostatic potential. The relevant contribution of the image particle comes from the region where the function $\mu$ vanishes, as a result, we can define the position of the image particle at $r=b$. The vanishing of $\mu$ causes a divergence in  Linet's potential $\psi(r,\theta)$, hence again close to the image particle $\psi(r,\theta)\approx V_c(r,\theta)$. The Copson potential behaves as (see \eqref{psiapproxip}) 
\begin{align}
\lim_{r\rightarrow b } \left(\mu\, V_c \right)&\approx \frac{q\, b/a}{r  \left(1+{r_s}/{4a}\right)^4}\quad , \quad |\boldsymbol{\nabla} \mu| \approx  \frac{(a-b)^{-1}}{\left(1+{r_s}/{4 a}\right)^2}\\
\boldsymbol{\nabla} V_c &\approx -\frac{V_c}{\mu} \boldsymbol{\nabla}  \mu  \quad \Rightarrow \quad |\boldsymbol{\nabla} V_c| \approx  \beta \, \left(\frac{V_c}{\Sigma}\right)^2 \nonumber\\
&\mbox{with } \quad 
\Sigma^2\equiv  \frac{\beta \, q(1-b/a)}{\left(1+{r_s}/{4a}\right)^2} \quad .
\end{align}
\color{black}
Following the same reasoning as before, the electrostatic potential reads
\begin{align}\label{phiip}
\phi_c (r,\theta)= -\frac{\Sigma^2}{V_c} \ {}_2F_1 \left[\frac14,\frac12,\frac54,-\left(\frac{\Sigma}{V_c}\right)^4\right]+\mbox{const.}
\end{align}

Equations \eqref{constitBI5} and \eqref{phiip} give the approximate solution of the electrostatic potential close to the charged particle outside the BH and its image particle inside the horizon respectively. 

In order to interpret the physical meaning of each contribution, let us analyze the divergent behavior of Linet's potential. Recall that in the case of a Maxwellian charged particle, this divergence can be absorbed in the renormalized rest mass of the particle~\cite{Bekenstein1999, Smith_Will1980}. To verify that the divergence has the  Coulombian form, it is useful to change to a coordinate system adapted to the particle. Hence, we define the coordinates $(t, \rho, \vartheta, \varphi)$
where 
%
the relation between  $(\rho, \vartheta)$ and $(r,\theta)$ is given by
\begin{align}\label{CoordTransf}
r \cos \theta \  \rightarrow  a+\rho \cos \vartheta\ \ ,\quad r \  \rightarrow  \sqrt{a^2+\rho^2+2a\rho \cos \vartheta}\quad .
\end{align}

The charged particle is located at $\rho=0$, hence it is reasonable to expand the potential using a Laurent series in powers of $\rho$
\begin{align}\label{psiattp}
\psi(\rho, \vartheta, \varphi)=&\frac{\left(1-r_s/4a\right)}{\left(1+r_s/4a\right)^3}\frac{q}{\rho}+\frac{q\, r_s}{2a^2\left(1+r_s/4a\right)^4}
\nonumber\\
&\quad +\frac{q\, r_s\left(2-r_s/4a\right)}{4a^2\left(1+r_s/4a\right)^4}\cos\vartheta+\mathcal{O}\left(\rho\right)
\end{align}

Clearly, for the limit $\rho\rightarrow0$, the dominant term (the first one) has a Coulomb potential behavior as $\psi \sim q/\rho$. The extra factors appear due to the change of variables. A factor $\left(1-r_s/4a\right)/\left(1+r_s/4a\right)$ is responsible for the redshift effect of the time component of the vector potential while an additional $\left(1+r_s/4a\right)^{-2}$ reflects the change of coordinate $\rho$ to proper distance. In terms of the coordinate system attached to the particle, the potential close to the particle must be isotropic with respect to the point $\rho=0$, hence we should average over all values of $\vartheta$ \cite{Bekenstein1999}. This cancels any contribution of the third term proportional to $\cos\vartheta$. The divergent Coulomb part can be absorbed by renormalizing the mass of the charged particle~\cite{Vilenkin1979,Smith_Will1980}. Then, for small values of $\rho$, the net contribution to the potential is the finite second term of \eqref{psiattp}.

In flat spacetime, the electrostatic potential of a small spherical charged body (or point-like particle) is given by the Coulomb potential. Thus, it is evident that the contribution of Eqn.\eqref{psiattp} is a superposition of the potential of the charged particle combined with the one coming from the image particle. Equivalently we can interpret it as the combined contribution of the charged particle and the polarization of the BH horizon. Indeed, Linet's potential is defined as a linear sum of Copson's solution with the contribution coming from the polarization of the BH horizon, i.e. $\psi(r,\theta)=V_c(r,\theta)+V_p(r,\theta)$ (see Eqn. \eqref{linetsol}). Using the coordinate transformation of
Eqn.\eqref{CoordTransf} and expanding in powers of $\rho$, we identify
\begin{align}
V_c(r,\theta)\rightarrow&
\frac{\left(1-r_s/4 a\right)}{\left(1+r_s/4 a\right)^3}\frac{q}{\rho}+\frac{q\, r_s\left(2-r_s/4 a\right)}{4 a^2\left(1+r_s/4 a\right)^4}\cos\vartheta +\mathcal{O}\left(\rho\right) \quad ,\nonumber\\
V_p(r,\theta) \rightarrow&\frac{q\, r_s}{2 a^2\left(1+r_s/4 a\right)^4}
+\mathcal{O}\left(\rho\right)\ .\label{SeriesVp}
\end{align}

\begin{figure}[t]
\includegraphics[width=0.47\textwidth]{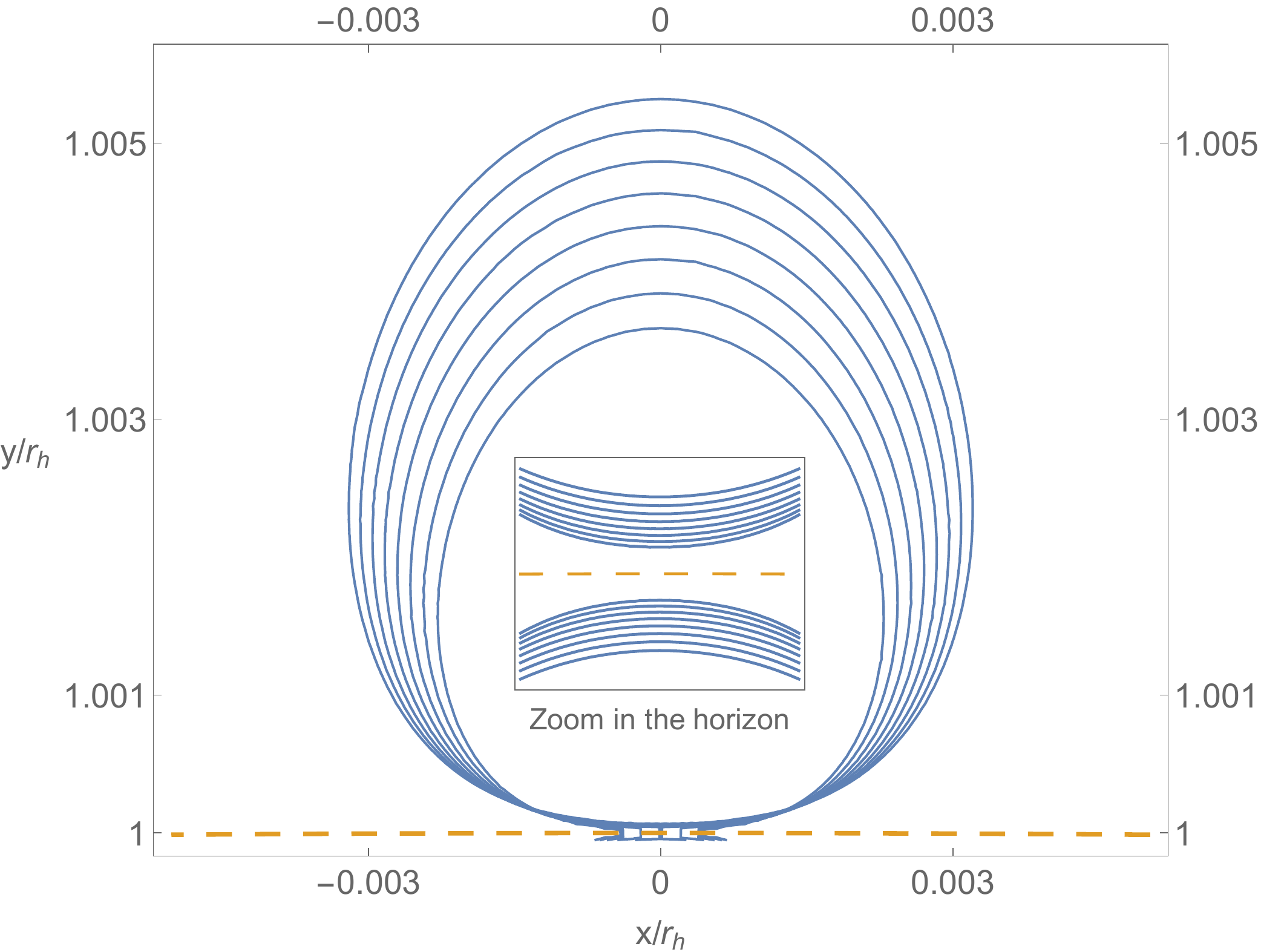}
\caption{In curved spacetime the BI region can be defined as the compact region inside which $|\boldsymbol{\nabla} \psi|\geq \beta$. We plot the BI region for a particle of charge $q=M$, $\beta=3M$. The horizon is located at $r_h=r_s/4$ that corresponds to the dashed line at the bottom. Each contour line defines the BI regions for different positions of the particle. We plot for $a/r_h = {1.0010, 1.0011, 1.0012, 1.0013, 1.0014, 1.0015, 1.0016, 1.0017}$. The box in the center shows a zoom of the region close to the horizon. Note that indeed close to the horizon the BI potential approaches
that of Linet. This is supported by the fact that none of the BI regions reach the horizon.}
\label{fig:BI_region_fig}
\end{figure}

Thus, the $V_c(r,\theta)$ gives the Coulomb like behavior and the finite contribution of \eqref{psiattp} comes from the polarization of the BH horizon. As already argued before, close to the horizon, the BI electrostatic potential approaches Linet's potential independently of the distance of the charged particle to the horizon. We can define the BI region, namely where the nonlinear corrections are important, as the interior of the boundary specified by $|\boldsymbol{\nabla}\psi|=\beta$. Figure~\ref{fig:BI_region_fig} shows how the BI region deforms as the charged particle approaches the horizon. Each curve delimits the BI region for different positions of the charged particle. The BI regions get squeezed and accumulate close to the horizon but never touch the horizon. This is consistent with the vanishing of the electric field at the BH horizon. Therefore, a BI charged particle produces the same polarization on the BH as a Maxwellian charged particle. As a consequence, the BI nonlinearities modify only the first term of Eqn.\eqref{psiattp} while keeping the contribution coming from the BH polarization unchanged.

\section{Lowering a Born-Infeld Body into a Schwarzschild Black Hole} \label{sec:5}

After Bekenstein's seminal paper~\cite{Bek81} proposing a universal bound for the entropy of a macroscopic body, a series of papers improved the bound by including the body's charge~\cite{Zaslavskii1992,Hod1999} and angular momentum~\cite{Hod99}. The optimal bound combining spin and charge of a macroscopic body was derived by Bekenstein and Mayo in~\cite{Bekenstein1999}. The bound follows from 
the application of the GSL~\cite{Bekenstein1974}, which states that the overall entropy of a physical system surrounded by a BH can never decrease. 

Since the entropy of a BH is determined by its horizon area, a bound on the object's entropy can be found by calculating the minimum increase of the BH area due to the object's infall. Bekenstein and Mayo~\cite{Bekenstein1999} showed that dropping a charged Maxwellian small body of radius $\mathcal{R}$, charge $q$ and proper mass $m$ causes an increment $\delta A$ of the area of the BH that obeys
\begin{equation}
\delta A\geq \frac{4\pi G}{c^4}\left(2mc^2 \mathcal{R}-q^2\right) \quad .
\label{deltaa}
\end{equation}

The entropy of the object cannot exceed the change in the area of the BH, since this would lead to the decrement of the entropy of the entire system (BH + object) after the object's assimilation. Thus, assuming GSL, we can set a bound on the entropy of the body as 
\begin{align}
\label{eq:Maxentropy}
S_{\rm BI}\leq & \frac{2\pi k_B}{\hbar c} \left[ mc^2\mathcal{R}- \frac{q^2}{2} \right]\quad .
\end{align} 

We shall generalize the above result by assuming that the field of the charged particle is governed by BI theory, in the presence of a Schwarzschild BH. The idea is to slowly lower the charged body from infinity to the region close to the horizon. Assuming an adiabatic process and keeping only first order corrections, the area of the BH should not change during the lowering process~\cite{Christodoulou1970, Bekenstein1999}. We can lower the body close to the horizon and then drop it into the BH. After a brief period of stabilization, the BH's entropy will increase due to the increase in its mass $M\rightarrow M+\mathcal{E}/c^{2}$ and the emergence of a net charge $q$.

The action associated with the motion $x^\mu(\tau)$ of a particle with rest mass $m$ and charge $q$ is~\cite{Carter1968, Misner1974}
\begin{equation}\label{AcP}
S=\int d\tau\left(mc\sqrt{g_{\mu\nu}\dot{x}^\mu\dot{x}^{\nu}}+\frac{q}{c}\dot{x}_\mu \hat{A}^{\mu}\right)\ ,
\end{equation}
where $\tau$ is the particle's proper time, a dot means time derivative with respect to $\tau$ and $\hat{A}^{\mu}$ is the background electromagnetic potential vector. The Schwarzschild BH has no net charge, hence does not directly contribute to $\hat{A}^{\mu}$. However, in curved spacetime the self-potential gives a nontrivial contribution to the energy as measured at infinity. Instead of reproducing Vilenkin~\cite{Vilenkin1979} and Smith and Will~\cite{Smith_Will1980} analysis, we shall follow Bekenstein and Mayo~\cite{Bekenstein1999} who argue that it  suffices to include a factor $\frac12$ in front of the self-potential, which instantiates the fact that part of the its energy comes from its own field and the rest from the background.

For a stationary spacetime, the timelike Killing vector $\xi^\mu=\left(1,0,0,0\right)$ defines the conserved quantity $\mathcal{E}=p_\mu \xi^\mu c$ , which is the energy measured at infinity. The momentum can be calculated directly from \eqref{AcP} by its definition $p_\mu=\p \mathcal{L}/\p \dot{x}^\mu$. Thus, it is straightforward to show that
\begin{align}
\mathcal{E}&= p_{\mu}\xi^{\mu}c=mc\dot{x}^{\beta}g_{0\beta}+{q\over2}A^0\ , \nonumber\\
&=mc^2
\frac{\left(1-r_s/4a\right)}{\left(1+r_s/4a\right)}
+{q\over2} \left(
\phi_c \left(a,\ 0\right)+\frac{q\, r_s}{2a^2\left(1+r_s/4 a\right)^4}\right)\ ,
\end{align}
where we have included the $1/2$ to account for the self-energy and the second line follows from combining the two contributions to the electrostatic potential (see \eqref{eq:phiapproximation} and \eqref{SeriesVp}). As shown in the previous section, the BI potential in the limit approaching the charged particle reads
\begin{equation*}
\phi_c (a,0)=\phi_{\rm BI}\sqrt{\frac{\left(1-r_s/4{a}\right)}{\left(1+r_s/4{a}\right)}}\quad .
\end{equation*}

When the particle is close to the horizon, we have $a\rightarrow r_s/4$, hence we can write $a-b=a(1-b/a)=a(1-r_s/4a)(1+r_s/4a)\approx2(a-r_s/4)$. The proper distance from the particle's center of mass to the horizon is~\cite{Bekenstein1999}
\[
l\equiv\int_{r_s/4}^{a}\dd r \sqrt{g_{rr}}\approx 4\left( a-r_s/4\right)\thinspace\Rightarrow 
\left\{\begin{array}{l}
a\approx \frac{r_s}{4}\left(1+\frac{l}{r_s} \right)\\
\\
{b}/{a}\approx 1-2{l}/{r_s}\\
\\
1+r_s/4a\approx2-l/r_s\\
\\
1-r_s/4 a\approx l/r_s
\end{array} \right.
\]
Hence, the energy of the particle as measured from infinity is
\begin{align}
\mathcal{E}&=\frac{1}{4r_s}\left[ 2mc^2l+q^2\left(1 +\sqrt{\frac{2lr_s}{\lambda_{\rm BI}^2}}\right)\right]+\mathcal{O}\left({l\over{M}}\right)^2,
\end{align}
where we have used the BI characteristic length scale $\lambda_{\rm BI}$ (see \eqref{BIMinkPotata}). The energy will be minimum when the proper distance from the particle to the horizon is equal to the particle's radius $\mathcal{R}$. Therefore, to leading order, the energy has a lower bound given by
\begin{align}
\mathcal{E}&\geq \frac{1}{4r_s}\left[ 2mc^2\mathcal{R}+q^2\left(1+\sqrt{\frac{2\mathcal{R}r_s}{\lambda_{\rm BI}^2}}\right)\right]
\quad .
\end{align}

After the particle is absorbed, the BH acquires a net charge $q$ and increases its mass $M$ to $M+\mathcal{E}/c^2$. The Reissner-Nordstr\"{o}m BH area is\footnote{Since $\beta$ is expected to be large, the expression for the radius of the external horizon of the Reissner-Nordstr\"{o}m BH can be used to calculate the change in $A$ instead of that of the BI black hole, see for instance
\cite{Sanchez2018}.}
$A=4\pi r_+^2=\pi r_s^2\left(1+\sqrt{1-4r_q^2/r_s^2}\right)^2$, where $r_q^2=G q^2/c^4$. To first order, we can approximate the area by $A=4\pi (r_s^2-2Gq^2/c^4)+\mathcal{O}(3)$. Thus, the change in the horizon area reads
\begin{align} \label{eq:changeHA}
\delta A&=\frac{8\pi G}{c^4}\left(2r_s \mathcal{E}-q^2\right)+\mathcal{O}(3)\nonumber\\
&\geq 4\pi
\left[ \frac{2Gm}{c^2}\mathcal{R}-\frac{Gq^2}{c^4}\left(1-\sqrt{\frac{2\mathcal{R}r_s}{\lambda_{\rm BI}^2}}\right)\right]\quad .
\end{align} 

The minimum change in the area is given by the equality in the above equation. We can restore the Bekenstein and Mayo result by dropping out the terms inversely proportional to $\lambda_{\rm BI}$. At first, this might seem a contradiction since Maxwell electrodynamics is recovered in the limit $\lambda_{\rm BI}\rightarrow0$. However, one must remember that Maxwell's electrodynamics gives a divergent contribution to the self-energy which is commonly absorbed in the mass term. In BI electrodynamics, the nonlinearity of the theory already gives a finite value for the self-energy that correspond to the above extra terms inversely proportional to $\lambda_{\rm BI}$. Thus, the divergent term with $\lambda_{\rm BI}$ can be absorbed in the renormalized mass of the charged particle, and the change of the area coincides with that calculated in \cite{Bekenstein1999}, namely $\delta A\geq 4\pi\left(2Gm\mathcal{R}/c^2- Gq^2/c^4 \right)$.

Note, however, there is a crucial difference between \eqref{deltaa} and \eqref{eq:changeHA} that allows one to associate the minimum change in area with an entropy bound. In the Maxwellian case, the change in the area of the BH depends only on the particle's properties. Therefore, even assuming GSL, one cannot promptly associate the quantity on the right hand side of \eqref{eq:changeHA} with the maximal value of the object's entropy. A straightforward solution is to use the trivial inequality $r_s\geq \mathcal{R}$ to eliminate the dependence on the BH mass. We exchange the strength of inequality \eqref{eq:changeHA} with a condition of the change in area of the BH that depends only on the particle's charge and mass. Thus, the maximal value of a charged particle's entropy satisfying the BI electrodynamics is given by
\begin{align} 
\label{eq:BIentropy}
S_{\rm BI}\leq & \frac{2\pi k_B}{\hbar c}  	\left[ 
mc^2\mathcal{R}+ \frac{q^2}{2}
\left(-1+\sqrt 2\frac{\mathcal {R}}{\lambda_{\rm BI}}\right)
\right]\quad .
\end{align} 
We recognize the first two terms on the right hand side as the bound previously established, hence BI electrodynamics raises the upper limit of the entropy of a macroscopic body. The characteristic length scale $\lambda_{\rm BI}$ is inversely proportional to the parameter $\beta$ , which is expected to be very large in order for BI electrodynamics to satisfy laboratory constraints. The size of the object has to be small compared to the BH radius but can be much larger than BI length, i.e. it is perfectly conceivable that $\mathcal{R}\gg \lambda_{\rm BI}$. Thus, instead of a small correction, the above extra term can greatly enlarge the entropy bound of a macroscopic body. Our result suggests that Bekenstein's inequalities heavily depend on the dynamical structure of the underlying theories.

\section{Conclusion} \label{sec:6}

The electrostatic potential for a charged particle satisfying Einstein-Maxwell dynamics in Schwarzschild has been derived long ago by Copson and then improved by Linet. We have generalized their results by proving theorem~\ref{thmNLEDSBH} of Sec.~\ref{sec:3}, that gives the electrostatic potential for an arbitrary NLED in Schwarzschild spacetime.
This is an interesting result that allows for the study of the self-energy of a charged particle and possibly for the radiation emitted by an accelerated BI charged particle, a problem that we intend to tackle in a future publication.

Bekenstein's inequality was deduced within BH thermodynamics and sets a bound on the entropy of a charged macroscopic body. It is envisaged as a universal relation between physical quantities and constants of nature, hence it should hold for arbitrary physical systems. That and other inequalities 
of the same kind
are derived via a gedanken experiment consisting of slowly lowering a small body close to the BH horizon and then calculating the minimum change of its area after the object's assimilation.

In the present work, we reanalyzed the same gedanken experiment but assuming that the charged object obeys BI electrodynamics. As in flat spacetime, BI gives a finite value for the electrostatic potential $\phi(x)$ and the modulus of the electric field $\mathbf{E}(x)$. Thus, there is no need to regularize the contribution coming from the particle's self-energy and the minimum change in the BH area gains a positive extra term. As a consequence, the entropy bound is augmented by a nonlinear contribution to the electromagnetic self-energy. This result is a definite proof that Bekenstein's inequalities depend on the underlying dynamical theory used to describe the physical system, and suggests that, contrary to previous claims
(see for instance
\cite{Bekenstein1990}
),  
nonlinear interactions may 
violate Bekenstein's entropy bound.

\section*{Acknowledgments}
The authors would like to thank and acknowledge financial support from the National Scientific and Technological Research Council (CNPq, Brazil).
SEPB would like to acknowledge support from UERJ and FAPERJ.

\appendix
\section{Mathematical expansions}
We shall analyse here the behavior and definition of some of the functions in the limit approaching the charged particle. The function $\mu({r},\theta)$ defined in \eqref{mufunct} is finite for any ${r}$ and $\theta\neq 0$. Furthermore, it never goes to zero outside the horizon and diverges only for $\theta=0$ and $r\rightarrow a$. Outside the horizon we need to expand it around the point $({r}={a}, \theta=0)$. Inside the horizon, $\mu$ goes to zero close to the point $({r}={b}, \theta=0)$, which also gives a divergence for the potential $\psi$. Thus we need to expand the potential around these two divergent points. Recall that 
\begin{align*}
\psi ({r},\theta)&=\frac{Q}{{r}\, \mu}\left(\frac{\mu+r_s/4{a}}{1+r_s/4{r}}\right)^2
\quad , \quad Q=q\left(1+r_s/4{a}\right)^{-2} \\
\mu({r},\theta)&=\sqrt{\frac{\left({r}-{b}\right)^2+2{b}{r}\left(1-\cos\theta\right)}{\left({r}-{a}\right)^2+2{a}{r}\left(1-\cos\theta\right)}}\quad , \qquad {b}\equiv \frac{r_s^2}{16{a}}\ .
\end{align*}
Let us start by calculating the relevant gradients. Direct calculation gives
\begin{align}
\frac{1}{\mu^3}\frac{\partial \mu}{\partial {r}}&
=({a}-{b})\frac{\left[({r}^2+{a}{b})(1-\cos\theta)-({r}-{a})({r}-{b})\right]}{\left({r}^2+{b}^2-2{b}{r}\cos\theta\right)^2}\label{gradmur1}\\
\mu\frac{\partial \mu}{\partial {r}}&
=({a}-{b})\frac{\left[({r}^2+{a}{b})(1-\cos\theta)-({r}-{a})({r}-{b})\right]}{\left({r}^2+{a}^2-2{a}{r}\cos\theta\right)^2}\label{gradmur2}
\end{align}
\begin{align}
\frac{1}{\mu^3}\frac{1}{{r}}\frac{\partial \mu}{\partial\theta}&=\frac{ ({a}-{b})({a}{b}-{r}^2)\sin\theta}{\left({r}^2+{b}^2-2{b}{r}\cos\theta\right)^2} \label{gradmutheta1}\\
\frac{\mu}{{r}}\frac{\partial \mu}{\partial\theta}&=
\frac{ ({a}-{b})({a}{b}-{r}^2)\sin\theta}{\left({r}^2+{a}^2-2{a}{r}\cos\theta\right)^2}\label{gradmutheta2}
\end{align}
\begin{align}
\frac{1}{\psi}\frac{\partial \psi}{\partial {r}}&=
-\frac{\left(1-r_s/4{r}\right)}{{r}\left(1+r_s/4{r}\right)}+\left(\frac{\mu-r_s/4{a}}{\mu+r_s/4{a}}\right)\frac{1}{\mu}\frac{\partial \mu}{\partial {r}}\label{gradpsi1}\\
\frac{1}{\psi}\frac{\partial \psi}{\partial\theta}&=\left(\frac{\mu-r_s/4{a}}{\mu+r_s/4{a}}\right)\frac{1}{\mu}\frac{\partial \mu}{\partial \theta}\label{gradpsi2}
\end{align}

There are three important regions to be considered: close to the charged particle, close to the image particle and close to the horizon. The horizon works as a ``conducting sphere'', which means that the electrostatic field vanishes and the potential has a constant value. Furthermore, the gradient of the potential is normal to the horizon surface. Indeed, taking the ${r}\rightarrow r_s/4$ limit we find
\begin{align}
\mu &\rightarrow\frac{r_s}{4{a}}\quad ,\quad 
\psi \rightarrow\frac{ q}{{a}\left(1+r_s/4{a}\right)^2}\quad ,\quad g^{11}\rightarrow-\frac{1}{16} \quad , \quad\nonumber\\
g^{22}&\rightarrow-\frac{1}{r_s^2} \quad , \quad \frac{\partial \mu}{\partial \theta}=\frac{\partial \psi}{\partial \theta}\rightarrow0\nonumber\\
|\boldsymbol{\nabla} \mu |&\rightarrow\frac{\left({a}^2- r_s^2/16\right)}{2{a}\left[2\left({a}- r_s/4\right)^2+{a}r_s\left(1-\cos \theta\right) \right]}\quad , \quad 
|\boldsymbol{\nabla} \psi |\rightarrow0
\end{align}
An important consequence is that close to the horizon the nonlinearities of the electrostatic potential are negligible and the potential approaches Linet's potential, i.e. in the limit ${r}\rightarrow r_s/4$ we have $\phi\approx\psi$. We can also take the limit of the charged particle approaching the horizon, i.e. ${a}\rightarrow r_s/4$, hence ${b}\rightarrow r_s/4={a}$. In this case $\mu \rightarrow 1$ and 
\begin{equation}
\lim_{{a}\rightarrow r_s/4}\ 
\psi ({r},\theta)=\frac{q}{{r}\left(1+r_s/4{r}\right)^2}\quad .
\end{equation}

Close to the charged particle $({r}\rightarrow {a}, \theta \rightarrow 0)$ the function $\mu$ diverges, hence we can expand it around this singular point as ${r}={a}+\delta$ and $\varepsilon=2(1-\cos \theta)$ giving 
\begin{align}
\mu^2(r,\theta)&=\frac{({a}-{b})^2+2\delta({a}-{b})+{b}\varepsilon({a}+\delta)}{\delta^2+{a}({a}+\delta)\varepsilon}\\
\frac{1}{\mu^3}\frac{\partial \mu}{\partial {r}}&=
\frac{({a}-{b})\left[
-2\delta({a}-{b})-2\delta^2+{a}({a}+{b})\varepsilon+2{a}\delta \varepsilon +\delta^2 \varepsilon
\right]}
{2\left[({a}-{b})^2+2\delta({a}-{b})+\delta^2+{a}{b}\varepsilon+{b}\delta\varepsilon\right]^2}
\nonumber\\
&=-\frac{\delta}{({a}-{b})^2}+\frac{3\delta^2}{({a}-{b})^3}+\frac{{a}({a}+{b})}{2({a}-{b})^3}\varepsilon+\mathcal{O}\left(\delta^3,\delta\varepsilon\right)
\label{dmudr}\\
\frac{1}{\mu^3}\frac{\partial \mu}{\partial\theta}&=
-\frac{({a}-{b})({a}+\delta)\left[{a}{b}-({a}+\delta)^2\right]\sqrt{\varepsilon(4-\varepsilon)}}
{2\left[({a}-{b})^2+2\delta({a}-{b})+\delta^2+{a}{b}\varepsilon+{b}\delta\varepsilon\right]^2}\nonumber\\
&=-\frac{{a}\sqrt{\varepsilon}}{({a}-{b})^2}\left({a}-\frac{({a}+{b})\delta}{({a}-{b})}\right)
+\mathcal{O}\left(\delta^2\sqrt{\varepsilon},\varepsilon^{3/2}\right)\label{dmudtheta2}
\end{align}

The derivatives of the potential read
\begin{align}
\frac{1}{\psi}\frac{\partial \psi}{\partial {r}}&=
-\frac{\left(1-r_s/4{r}\right)}{{r}\left(1+r_s/4{r}\right)}+\left(\frac{\mu-r_s/4{a}}{\mu+r_s/4{a}}\right)\frac{1}{\mu}\frac{\partial \mu}{\partial {r}}\nonumber\\
&=\frac{\partial}{\partial {r}}\log\left(\frac{\mu}{{r}(1+r_s/4{r})^2}\right)
+\mathcal{O}\left(\delta,\varepsilon\right)\label{dpsidr}\\
\frac{1}{\psi}\frac{\partial \psi}{\partial\theta}&=\left(\frac{\mu-r_s/4{a}}{\mu+r_s/4{a}}\right)\frac{1}{\mu}\frac{\partial \mu}{\partial \theta}\nonumber\\
&=\frac{\partial }{\partial \theta}\log \mu+\mathcal{O}\left(\delta,\varepsilon\right)\label{dpsidtheta}
\end{align}

Therefore, in the limit approaching the real charged particle we have $\psi \approx V_c$ and 
\begin{align}
\frac{V_c}{\mu} &\approx \frac{q}{{r}  \left(1+{r_s}/{4{a}}\right)^4}
\quad , \quad |\boldsymbol{\nabla} \mu| \approx  \frac{ \mu^2({a}-{b})^{-1}}{\left(1+{r_s}/{4{a}}\right)^2}\\
\boldsymbol{\nabla}  V_c &\approx \frac{V_c}{\mu} \boldsymbol{\nabla}  \mu  \quad \Rightarrow \quad |\boldsymbol{\nabla} V_c| \approx  \beta \, \left(\frac{V_c}{\Sigma}\right)^2 \nonumber\\
&\mbox{with } \quad
\Sigma^2\equiv  \frac{\beta \, q(1-{b}/{a})}{\left(1+{r_s}/{4{a}}\right)^2}\label{psiapproxcp}
\end{align}

We can do a similar analysis close to the image particle where the function $\mu$ goes to zero. Approaching the image particle $({r}\rightarrow{b}, \theta \rightarrow 0)$ we can expand the functions using ${r}={b}+\delta$ and $\varepsilon=2(1-\cos \theta)$. Thus we have
\begin{align}
\mu^2(r,\theta)&=\frac{\delta^2+{b}({b}+\delta)\varepsilon}{({a}-{b})^2-2\delta({a}-{b})+\delta^2+{a}{b}\varepsilon}\\
\mu\frac{\partial \mu}{\partial {r}}&=
({a}-{b})
\frac{\left[({b}+\delta)^2+{a}{b}\right]\varepsilon-2\delta(\delta+{b}-{a})}{2\left[(\delta+{b}-{a})^2+{a}(\delta+{b})\varepsilon\right]^2}
\nonumber\\
&=
\frac{\delta}{({a}-{b})^2}+\frac{3\delta^2}{({a}-{b})^3}+\frac{{b}({a}+{b})}{2({a}-{b})^3}\varepsilon+\mathcal{O}\left(\delta^3,\delta\varepsilon\right)
\label{dmudr_2}
\end{align}
\begin{align}
\mu\frac{\partial \mu}{\partial\theta}
&=\frac{({a}-{b}) }{2}\frac{\left[{a}{b}-({b}+\delta)^2\right]\sqrt{\varepsilon(4-\varepsilon)}}{2\left[(\delta+{b}-{a})^2+{a}(\delta+{b})\varepsilon\right]^2}
\nonumber\\
&=
\frac{{b}\sqrt{\varepsilon}}{({a}-{b})^2}\left({b}+\frac{({a}+{b})\delta}{({a}-{b})}\right)
+\mathcal{O}\left(\delta^2\sqrt{\varepsilon},\varepsilon^{3/2}\right)
\label{dmudtheta2_2}
\end{align}

Similarly to \eqref{psiapproxcp}, close to the imagine particle $(\mu\rightarrow0)$, we have
\begin{align}
\mu\, V_c &\approx \frac{q\left(r_s/4{a}\right)^4}{{r} \left(1+{r_s}/{4{a}}\right)^4}\quad , \quad |\boldsymbol{\nabla} \mu| \approx  \frac{({a}-{b})^{-1}}{\left(1+{r_s}/{4{b}}\right)^2}\\
\boldsymbol{\nabla}  V_c &\approx -\frac{V_c}{\mu} \boldsymbol{\nabla}  \mu  \quad \Rightarrow \quad |\boldsymbol{\nabla} V_c| \approx  \beta \, \left(\frac{V_c}{\Sigma}\right)^2 \nonumber\\
&\mbox{with } \quad 
\Sigma^2\equiv  \frac{\beta \, q(1-{b}/{a})}{\ \left(1+{r_s}/{4{a}}\right)^2}\ .\label{psiapproxip}
\end{align}


\bibliography{Bibliografia}

\end{document}